\documentclass[journal]{IEEEtran}
\IEEEoverridecommandlockouts
\usepackage{cite}
\usepackage{amsmath,amssymb,amsfonts,mathtools,bm,amsthm}
\usepackage{algorithmic}
\usepackage{graphicx}
\usepackage{textcomp}
\usepackage{xcolor}
\usepackage{placeins}
\usepackage{tikz}\usetikzlibrary{positioning}
\usepackage{makecell}
\usetikzlibrary{calc}
\usetikzlibrary{arrows.meta}
\usetikzlibrary{arrows.meta, fit, backgrounds}
\usetikzlibrary{backgrounds, shapes.misc}
\usetikzlibrary{shapes.geometric}
\allowdisplaybreaks
\usepackage{soul}
\usepackage{stfloats}
\usepackage{tabularx}
\usepackage{booktabs}
\usepackage{array}
\usepackage{hyperref}
\usepackage{url}
\usepackage{graphicx} 
\usepackage{diagbox}
\usepackage[ruled,vlined,linesnumbered]{algorithm2e} 
\usepackage{xcolor} 
\usepackage[ruled,vlined]{algorithm2e}
\def\BibTeX{{\rm B\kern-.05em{\sc i\kern-.025em b}\kern-.08em
 T\kern-.1667em\lower.7ex\hbox{E}\kern-.125emX}}
\usepackage{standalone}
\usepackage{tikz}\usetikzlibrary{calc, shapes, arrows, automata, positioning, patterns}

\begin{document}

\title{Graph approach for observability analysis in power system dynamic state estimation}
\author{Akhila~Kandivalasa,~\IEEEmembership{Student~Member,~IEEE,}~Marcos~Netto,~\IEEEmembership{Senior~Member,~IEEE}
\thanks{This work is supported by the National Science Foundation under Grant 2328241. The authors are with the Department of Electrical and Computer Engineering, New Jersey Institute of Technology, Newark, NJ 07102, USA.}
}

\maketitle

\begin{abstract}
The proposed approach yields a numerical method that provably executes in linear time with respect to the number of nodes and edges in a graph. The graph, constructed from the power system model, requires only knowledge of the dependencies between state-to-state and output-to-state variables within a state-space framework. While \emph{graph-based} observability analysis methods exist for power system static-state estimation, the approach presented here is the first for dynamic-state estimation (DSE). We examine decentralized and centralized DSE scenarios and compare our findings with a well-established, albeit non-scalable, observability analysis method in the literature. When compared to the latter in a centralized DSE setting, our method reduced computation time by \texttt{1440x}.
\end{abstract}


\section{Introduction}
\emph{Observability} refers to the ability to determine a system's state from measurements. While well-understood for power system static state estimation \cite{Abur2004}, observability analysis is an active research area within dynamic state estimation (DSE) \cite{Zhao2019}, with several challenges still to be addressed.

A grand challenge stems from the computational complexity of the numerical methods used for observability analysis in dynamic state estimation. The mainstream approach relies on Lie differentiation \cite{Hermann_Krener_1977}. In what follows, for simplicity, we will refer to the class of methods based on Lie derivatives as the $\mathcal{L}$ \emph{approach}. The $\mathcal{L}$ approach does not scale effectively \cite{deJager1995}, as noted by power system researchers \cite{Rouhani2017}. Indeed, we confirmed this limitation. Our implementation of the Western System Coordinating Council (WSCC) test case (Example 7.1 in \cite{Sauer1997}) takes about 2 hours to run in \texttt{MATLAB}. Note that this test case has only twenty-one state variables. Although implementation details and computing power may differ, it is indisputable that the $\mathcal{L}$ approach is impractical for power system operators, as their systems often have thousands of state variables.

The empirical observability Gramian (EOG) \cite{Krener2009} might serve as an alternative to the $\mathcal{L}$ approach. Unlike the $\mathcal{L}$ approach, which relies on \emph{symbolic} computation of Lie derivatives, the EOG provides a way to evaluate observability \emph{numerically}. One constructs the EOG by (i) applying small perturbations to the system’s initial states, (ii) simulating the corresponding output trajectories over a finite horizon, and (iii) assembling these sensitivities into a symmetric matrix. This symmetric matrix corresponds to the Fisher information matrix, and its determinant (the product of its eigenvalues) measures the information the outputs carry about the initial state, and can thus be used to assess observability \cite{Qi2015}. However, this reliance on perturbing the system and collecting data---which makes it \emph{empirical}---introduces a key disadvantage. This approach requires recomputation and additional simulations across many operating conditions, time windows, and outputs \cite{Krener2009}. Repeated numerical simulations make the process computationally intensive, limiting its scalability to large-scale power grids. Hence, this approach does not solve the challenge of computational complexity.

Since the EOG and $\mathcal{L}$ approaches do not scale, an alternative could be to perform an observability analysis on a linear approximation of the original nonlinear system. However, relying on a linear approximation might lead to incorrect conclusions, as Rouhani and Abur \cite{Rouhani2017} show using a \emph{linearized} fourth-order synchronous machine model. They create an ideal scenario where all state variables are assumed to be measured and a second scenario where only a single reactive power measurement is available. One would expect the ideal scenario to yield stronger observability, since all state variables are directly measured. Yet, because of linearization, observability analysis favors the second (non-ideal) scenario \cite{Rouhani2017}.


This paper departs from the EOG and $\mathcal{L}$ approaches, and does not rely on linearization. We introduce a method for analyzing observability that uses a directed graph (digraph) constructed from the nonlinear differential equations \cite{Liu2013} that model power systems. From now on, we refer to the proposed method as the \emph{graph approach}. We show that the graph approach produces results equivalent to those of the $\mathcal{L}$ approach but needs significantly less computational effort. Specifically, for the WSCC test case with the same conditions used to evaluate the $\mathcal{L}$ approach, it takes less than 5 seconds to check for observability using the proposed method. At the core of this method is Kosaraju-Sharir's algorithm, which finds all \emph{strongly-connected components} (SCCs)---defined in Section \ref{sec:results} and central to the proposed method---in a digraph and operates in linear time relative to the number of nodes and edges \cite{SHARIR198167}. Therefore, the proposed method can scale to power grids of any size. While there are numerical and graph-based (topological) observability analysis methods for power system static state estimation \cite{Netto2022}, this is the first time a graph-based method has been proposed for observability analysis in power system \emph{dynamic state estimation} \cite{Zhao2019}.

This paper proceeds as follows. Section \ref{sec:prelim} briefly presents the $\mathcal{L}$ approach to observability analysis. Section \ref{sec:graph} introduces the proposed graph approach. Section \ref{sec:results} demonstrates the graph approach for both decentralized and centralized dynamic state estimation. Section \ref{sec:conclusion} summarizes the key findings and outlines future work.

\section{The \texorpdfstring{$\mathcal{L}$}{L} approach to observability analysis}\label{sec:prelim}
Consider a nonlinear dynamical system,
\vspace{-.1cm}
\begin{equation}\label{nonlinearsystem1new}
\bm{\dot{x}}(t) = \bm{f}(\bm{x}(t), \bm{u}(t)), \quad \bm{y}(t) = \bm{h}(\bm{x}(t))
\end{equation}

\vspace{-.1cm}
\noindent
evolving on a smooth manifold $\mathcal{M} \subseteq \mathbb{R}^n$, where the state $\bm{x} \in \mathcal{M}$, the input $\bm{u} \in \mathbb{R}^m$, and the output $\bm{y} \in \mathbb{R}^p$; $\bm{f}$ defines the system dynamics, and $\bm{h}$ relates measured outputs with state variables. The notion of indistinguishability \cite{Hermann_Krener_1977} provides the basis for defining observability in dynamical systems. 

\noindent
\textbf{Definition 1.} Let $\bm{x}(0)\eqqcolon \bm{x}_{0}$ denote a possible state of \eqref{nonlinearsystem1new} at $t=0$. Two states $\bm{x}_0,\, \bm{x}_0^{\prime} \in \mathcal{M}$ are \emph{indistinguishable} if for every admissible input $\bm{u}(t)$, the resulting output trajectories are identical over a finite time interval, i.e.,
$\bm{h}(\bm{x}(t; \bm{x}_0, \bm{u})) = \bm{h}(\bm{x}(t; \bm{x}_0^{\prime}, \bm{u}))$ for all $\bm{u}(t)$, $t \in [t_0, t_1]$. 

\noindent
The set of all such indistinguishable states from $\bm{x}_0$ is denoted as $\mathcal{I}(\bm{x}_0) = \{ \bm{x}_0,\,\bm{x}_0^{\prime}\}$. 

\noindent
\textbf{Definition 2.} A system \eqref{nonlinearsystem1new} is \textit{observable at $\bm{x}_0$} if the only state indistinguishable from $\bm{x}_0$ is $\bm{x}_0$ itself, that is, $\mathcal{I}(\bm{x}_0) = \{ \bm{x}_0 \}$. 

While precise, Definition 2 is global in nature and requires knowledge of the entire space of input--output trajectories over the considered time interval. Such a condition is challenging to verify numerically for nonlinear systems. So, Hermann and Krener \cite{Hermann_Krener_1977} proceed by defining indistinguishability locally over a small neighborhood $\mathcal{V}$ in an open neighborhood $\mathcal{U}$ of $\bm{x}$, $\mathcal{V}\subset \mathcal{U}$, from which another definition for observability emerges.

\noindent
\textbf{Definition 3.} A system \eqref{nonlinearsystem1new} is \emph{locally weakly observable at $\bm{x}_0$} if $\mathcal{I}_{\mathcal{V}}(\bm{x}_0) = \{ \bm{x}_0 \}$, and is \emph{locally weakly observable} if it is so at every $\bm{x}\in\mathcal{M}$.

Simply put, based on Definition 3, a system is observable if no other state produces the same input--output trajectory within a small neighborhood of $\bm{x}$. Definition 3 is amenable to a numerical test to check for observability---the $\mathcal{L}$ approach, which we introduce next. We refer the interested reader to \cite{Hermann_Krener_1977} for a more formal treatment and further details on observability definitions. Now, consider a map
\vspace{-.1cm}
\begin{equation}\label{outputmap}
\mathcal{G} \; :\; \bm{x}\mapsto
\begin{bmatrix}
\bm{y} & \bm{y}^{(1)} & \bm{y}^{(2)} & \cdots & \bm{y}^{(d)}
\end{bmatrix}^\top
\end{equation}

\vspace{-.1cm}
\noindent
where $\bm{y}^{(k)}$ denotes the $k$th-order derivative of $\bm{y}$, $k\in\mathbb{N}$. If $\mathcal{G}$ is invertible for some $d\le(n-1)$, then one can reconstruct the state from measurements; hence, the system is observable. If we consider Definition 3, then the Implicit Function Theorem provides a sufficient condition for local invertibility. The map $\mathcal{G}$ is locally invertible at $\bm{x}_{0}$ if the Jacobian
\vspace{-.1cm}
\begin{equation}
\mathcal{O} \coloneqq \frac{\partial\mathcal{G}}{\partial\bm{x}}\bigg|_{\bm{x}=\bm{x}_{0}}
\end{equation}

\vspace{-.1cm}
\noindent
referred to as the \emph{observability matrix} has full rank, i.e., if
\vspace{-.1cm}
\begin{equation}
\operatorname{rank}\left(\frac{\partial\mathcal{G}}{\partial\bm{x}}\bigg|_{\bm{x}=\bm{x}_{0}}\right)=n. \label{eq:rank}
\end{equation}

\vspace{-.1cm}
If condition \eqref{eq:rank} holds, \eqref{nonlinearsystem1new} is locally weakly observable. 

\noindent
Rouhani and Abur \cite{Rouhani2017} apply condition \eqref{eq:rank}. In what follows, for simplicity, a system satisfying condition \eqref{eq:rank} is said \emph{observable}, although the $\mathcal{L}$ approach derives from Definition 3. Now, the Lie derivatives of $\bm{y}$ along the vector field $\bm{f}$
\vspace{-.1cm}
\begin{equation}\label{lie}
\mathcal{L}_{\bm{f}}\bm{y} \coloneqq \frac{\partial \bm{h}(\bm{x})}{\partial x_1} \dot{x}_1 + \frac{\partial \bm{h}(\bm{x})}{\partial x_2} \dot{x}_2 + \dots + \frac{\partial \bm{h}(\bm{x})}{\partial x_n} \dot{x}_n 
\end{equation}

\vspace{-.1cm}
\noindent
measure how outputs $\{y_{1},...,y_{p}\}$ in $\bm{y}$ change along the vector field $\bm{f}$. Higher-order Lie derivatives are given by
\vspace{-.1cm}
\begin{equation}\label{eq:liehighernew}
\mathcal{L}_{\bm{f}}^{k}\bm{y} = \frac{\partial \mathcal{L}_{\bm{f}}^{k-1} \bm{h}(\bm{x})}{\partial \bm{x}} \bm{f}, \quad 1\le k \le n,
\end{equation}

\vspace{-.1cm}
\noindent
and the zeroth order $\mathcal{L}_{\bm{f}}^{0}\bm{y} \coloneqq \bm{h}(\bm{x})$. With these definitions, the map \eqref{outputmap} becomes
\vspace{-.15cm}
\begin{equation}
\mathbf{g} =
\left[\arraycolsep=2.9pt \begin{array}{cccccc}
\mathcal{L}_{\bm{f}}^{0}y_1 &
\cdots &
\mathcal{L}_{\bm{f}}^{n-1}y_1 &
\;\mathcal{L}_{\bm{f}}^{0}y_2 &
\cdots &
\mathcal{L}_{\bm{f}}^{n-1}y_p
\end{array}\right]^\top
\end{equation}

\vspace{-.1cm}
\noindent
and the observability matrix $\mathcal{O} = \frac{\partial\mathbf{g}}{\partial\bm{x}}\big|_{\bm{x}=\bm{x}_{0}}$.

\vspace{.1cm}
\noindent
\textbf{Example 1.}
Let a nonlinear dynamical system be
\vspace{-.1cm}
\begin{equation}
\dot{x}_1 = \sin(x_2) - x_1 + x_4,\; 
\dot{x}_2 = x_3,\;
\dot{x}_3 = x_1^2,\;
\dot{x}_4 = x_4 \label{eq:nonlinear_exampl}
\end{equation}

\vspace{-.1cm}
\noindent
and consider three measurement models: (i) $y=h(\bm{x})= x_2$ (ii) $y^{\prime}=h(\bm{x})= x_2+\sin(x_1)$ and (iii) $y'^{\prime}=h(\bm{x})= x_4$. For case (i), $\mathcal{L}_{\bm{f}}^0{y} = x_2$, and
\vspace{-.1cm}
\begin{align*}
\mathcal{L}_{\bm{f}}^1{y}
&=\frac{\partial\mathcal{L}_{\bm{f}}^0{y}}{\partial \bm{x}} \!\bm{f}
=\frac{\partial {x_2}}{\partial x_1}\dot x_1
+ \frac{\partial {x_2}}{\partial x_2}\dot x_2
+ \frac{\partial {x_2}}{\partial x_3}\dot x_3
+ \frac{\partial {x_2}}{\partial x_4}\dot x_4 \\[0pt]
&= 0\cdot \dot x_1 + 1\cdot \dot x_2 + 0\cdot \dot x_3 + 0\cdot \dot x_4 = \dot x_2 = x_3 \\
\mathcal{L}_{\bm{f}}^2{y}
&=\frac{\partial\mathcal{L}_{\bm{f}}^1{y}}{\partial \bm{x}} \!\bm{f}
=\frac{\partial {x_3}}{\partial x_1}\dot x_1
+ \frac{\partial {x_3}}{\partial x_2}\dot x_2
+ \frac{\partial {x_3}}{\partial x_3}\dot x_3
+ \frac{\partial {x_3}}{\partial x_4}\dot x_4 \\[0pt]
&= 0\cdot \dot x_1 + 0\cdot \dot x_2 + 1\cdot \dot x_3 + 0\cdot \dot x_4 = \dot x_3 = x_1^2 \\
\mathcal{L}_{\bm{f}}^3{y}
&=\frac{\partial\mathcal{L}_{\bm{f}}^2{y}}{\partial \bm{x}} \!\bm{f}
=\frac{\partial {x_1^2}}{\partial x_1}\dot x_1
+ \frac{\partial {x_1^2}}{\partial x_2}\dot x_2
+ \frac{\partial {x_1^2}}{\partial x_3}\dot x_3
+ \frac{\partial {x_1^2}}{\partial x_4}\dot x_4 \\[0pt]
&= 2x_1\dot x_1 = 2x_1\big(\sin x_2 - x_1 + x_4\big)
\end{align*}

\vspace{-.1cm}
\noindent
Note that applying higher-order Lie derivatives is ``like peeling off layers to reveal hidden dependencies between outputs and state variables, including those not explicit in the original output function." The observability matrix for case (i) is
\vspace{-.1cm}
\begin{equation*}
\mathbf{O}_{1}(\bm{x})=
\begin{bmatrix}
0 & 1 & 0 & 0\\
0 & 0 & 1 & 0\\
2x_1 & 0 & 0 & 0\\
{2\sin x_2 - 4x_1+2x_4} & 2x_1\cos x_2 & 0 & 2x_1
\end{bmatrix}.
\end{equation*}

\vspace{-.1cm}
\noindent
Under generic conditions such as $x_1 \ne 0$, the rows of $\mathbf{O}_{1}(\bm{x})$ are linearly independent, and the matrix attains full rank. The system is, therefore, observable. The observability matrix for case (ii) is
\vspace{-.3cm}
\begin{equation*}
\mathbf{O}_2(x) =
\begin{bmatrix}
\cos x_1 & 1 & 0 & 0 \\[0pt]
\star_{2,1}(\bm{x}) & \star_{2,2}(\bm{x}) & \star_{2,3}(\bm{x}) & \star_{2,4}(\bm{x}) \\[0pt]
\star_{3,1}(\bm{x}) & \star_{3,2}(\bm{x}) & \star_{3,3}(\bm{x}) & \star_{3,4}(\bm{x}) \\[0pt]
\star_{4,1}(\bm{x}) & \star_{4,2}(\bm{x}) & \star_{3,4}(\bm{x}) & \star_{4,4}(\bm{x}) \\[0pt]
\end{bmatrix}.
\label{eq:O2}
\end{equation*}

\vspace{-.1cm}
\noindent
Case (ii) yields long symbolic expressions. For instance,
\begin{align*}
\star_{3,2}(\bm{x}) =
 -\cos(x_2) (\cos(x_1) + \sin(x_1)(x_4 - x_1 + \sin(x_2))\\
 - x_3 \cos(x_1)\sin(x_2) - \cos(x_2)\sin(x_1)(x_4 - x_1 + \sin(x_2))
\end{align*}

\vspace{-.1cm}
\noindent
For simplicity, we denote those expressions by $\star_{\text{row},\text{col}}(\bm{x})$ in $\mathbf{O}_2$. It can be verified that $\mathbf{O}_2$ is generically of full rank, hence the system is observable. For case (iii), $\mathcal{L}_{\bm{f}}^0{y^{\prime}} = x_4$, and

\vspace{-.3cm}
\begin{minipage}{0.4\linewidth}
\begin{align*}
\mathcal{L}_{\bm{f}}^1{y^{\prime}} &= 5x_4 \\
\mathcal{L}_{\bm{f}}^2{y^{\prime}} &= 25x_4 \\
\mathcal{L}_{\bm{f}}^3{y^{\prime}} &= 125x_4 \\
\end{align*}
\end{minipage}
\begin{minipage}{0.49\linewidth}
\centering
$
\mathbf{O}_{3}(\bm{x})=
\begin{bmatrix}
0 & 0 & 0 & 1\\
0 & 0 & 0 & 5\\
0 & 0 & 0 & 25\\
0 & 0 & 0 & 125
\end{bmatrix}.
$
\end{minipage}

\vspace{-.2cm}
\noindent
Clearly, $\mathbf{O}_{3}$ is rank deficient, and the system is not observable.\qed

In general, dim$(\mathcal{O})=np\times{n}$. However, the rank condition \eqref{eq:rank} may be satisfied before computing all Lie derivatives up to order $n-1$. In such cases, the hidden dependencies between outputs and state variables are unveiled at $k < n$. Therefore, the dimension of $\mathcal{O}$ is more accurately given by $(k+1)p \times n$, where $(k+1) \leq n$ \cite{Krener2009}. Even when the rank condition is satisfied at a lower order $k < n$, the intermediate symbolic expressions associated with the Lie derivatives rapidly expand in size. This is true even for the simple system \eqref{eq:nonlinear_exampl}, as evidenced from the structure of $\mathbf{O}_2$. This makes the $\mathcal{L}$ approach computationally demanding, and applying it to multi-machine power systems is infeasible. Rouhani and Abur \cite{Rouhani2017} apply the $\mathcal{L}$ approach to a fourth-order synchronous generator model with an IEEE type-1 exciter, a system with seven state variables. Section \ref{sec:results} of this paper demonstrates the proposed graph approach using the same decentralized model in \cite{Rouhani2017}. It further presents results for centralized dynamic state estimation.

\section{The graph approach to observability analysis}\label{sec:graph}


Let a digraph built from \eqref{nonlinearsystem1new} be given by $\mathcal{D} = (\mathcal{V}, \mathcal{E})$, where $\mathcal{V}$ is a set of nodes, and $\mathcal{E}$ is a binary relation on $\mathcal{V}$. The elements of $\mathcal{E}$ are called edges. Each node corresponds to a state variable $x_i\in\mathcal{V}$, and a directed edge $x_j \!\to\! x_i$ exists whenever the differential equation $\dot{x}_i$ explicitly depends on $x_j$, i.e., $\partial \dot{x}_i/\partial x_j \neq 0$. As defined, $\mathcal{D}$ may include self-loops, i.e., $x_i \!\to\! x_i$ edges. The set of all such dependencies can be compactly represented in an adjacency matrix with entries $a_{ij} = 1$ if $(x_j \!\to\! x_i) \in \mathcal{E}$ and $0$ otherwise.


\noindent
\textbf{Definition 4.} A \emph{path} $x_{j}$--$x_{i}$ is a sequence $\left(\varepsilon_{0},\varepsilon_{1},\dots,\varepsilon_{k}\right)$ with $\varepsilon_{0}=x_{j}$ and $\varepsilon_{k}=x_{i}$, and $\left(\varepsilon_{\ell-1},\varepsilon_{\ell}\right)\in\mathcal{E}\;\forall\ell\in\{\ell\}_{\ell=1}^{k}$.

\noindent
\textbf{Definition 5.} A node $x_i$ is \emph{reachable} from $x_j$ if there is a path from $x_j$ to $x_i$.

\noindent
\textbf{Definition 6.} A digraph $\mathcal{D}$ is \emph{strongly connected} if every two nodes are reachable from each other.

\noindent
\textbf{Definition 7.} A \emph{subgraph} $\mathcal{D}_{\text{sub}}=\left(\mathcal{V}_{\text{sub}}, \mathcal{E}_{\text{sub}}\right)$ of $\mathcal{D}$ satisfies $\mathcal{V}_{\text{sub}}\subseteq\mathcal{V}$ and $\mathcal{E}_{\text{sub}}\subseteq\mathcal{E}$, with $\mathcal{E}_{\text{sub}}$ containing ordered pairs.

\noindent
\textbf{Definition 8.} \emph{Strongly connected components} (SCCs) are the largest subgraphs, ordered by number of nodes, such that each component is a strongly connected graph.

\noindent
\textbf{Definition 9.} A \emph{root strongly connected component} (root SCC) is an SCC with no incoming edges.

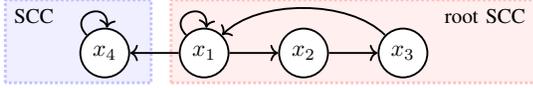
\begin{figure}[!t]\centering
\begin{tikzpicture}[node distance={15mm}, thick, main/.style = {draw, circle}] 
\filldraw[densely dotted, color=red!60, fill=red!15, very thick, opacity=0.4] (-.5,-0.45) rectangle (5,.8);
\filldraw[densely dotted, color=blue!60, fill=blue!15, very thick, opacity=0.4] (-3,-0.45) rectangle (-.8,.8);
\draw (3.5,.8) node[anchor=north west] {\small root SCC};
\draw (-3,.8) node[anchor=north west] {\small SCC};
\node[main] (1) [            fill=white] {$x_{1}$};
\node[main] (2) [ left of=1, fill=white] {$x_{4}$};
\node[main] (3) [right of=1, fill=white] {$x_{2}$};
\node[main] (4) [right of=3, fill=white] {$x_{3}$};
%
\draw[->] (1) to [out=135,in=85,looseness=3.5] (1);
\draw[->] (2) to [out=135,in=85,looseness=3.5] (2);
\draw[->] (4) to [out=135,in=45,looseness=0.8] (1);
\draw[->] [color=black] (1) -- (2);
\draw[->] [color=black] (1) -- (3);
\draw[->] [color=black] (3) -- (4);
\end{tikzpicture}
\vspace{-3mm}
\caption{Digraph of \eqref{eq:nonlinear_exampl}.}
\label{fig:digraphexample}
\end{figure}

\noindent
\textbf{Example 2}. The digraph of \eqref{eq:nonlinear_exampl} (Fig. \ref{fig:digraphexample}) has six directed edges: $x_1\!\to\!x_1$, $x_1\!\to\!x_2$, $x_1\!\to\!x_4$, $x_2\!\to\!x_3$, $x_3\!\to\!x_1$, and $x_4\!\to\!x_4$. Nodes $x_1$, $x_2$, and $x_3$ form an SCC, since one can reach every node from every other node. Furthermore, since there are no incoming edges to this SCC, it is also a root SCC.\qed

Root SCCs act as information sources within $\mathcal{D}$ because all nodes in $\mathcal{D}$ can be traced back to at least one node within a root SCC; however, the reverse is not true.

\noindent
\textbf{Proposition 1.} A dynamical system \eqref{nonlinearsystem1new} is termed \emph{structurally observable} if, for a digraph $\mathcal{D}$ built from \eqref{nonlinearsystem1new}, at least one node from every root SCC is measured \cite{Liu2013}.

\begin{algorithm}[!t]
\caption{\textcolor{black}{Extraction of SCCs and Root SCCs from Adjacency Matrix}}
\label{algo:scc-mat}
\SetAlgoLined
\SetKwInOut{Require}{Require}
\SetKwInOut{Ensure}{Ensure}

\Require{Adjacency matrix $A = [a_{ij}] \in \{0,1\}^{n \times n}$}
\Ensure{SCCs, $\mathcal{C} = \{C_1,\dots,C_m\}$, and root SCCs, $\mathcal{R}$}

\textbf{Step 1: Compute finishing order of nodes}\footnotetext[1]{Nodes are recursively explored until no further connected nodes remain; each node is then pushed to the stack, so the last finished node is on top}\;
visit $\leftarrow [0,\dots,0]_{1\times n}$, stack $\leftarrow$ empty\;
\For{$i \gets 1$ \KwTo $n$}{
    \If{\emph{visit}$[i]=0$}{
        [visit, stack] $\leftarrow$ FillOrder($i$, $A$, visit, stack)\;
    }
}
\textbf{Procedure} FillOrder($u$, $A$, visit, stack):\\
    visit$[u]\leftarrow1$\;
    \For{$v\gets1$ \KwTo $n$}{
        \If{$A_{uv}=1$ and \emph{visit}$[v]=0$}{[visit, stack] $\leftarrow$ FillOrder($v$, $A$, visit, stack)\;
    }
    }
    append $u$ to stack\;

\textbf{Step 2: Identify SCCs on transposed graph}\;
$A^\top$ = transpose(A); visit $\leftarrow [0,\dots,0]_{1\times n}$; $\mathcal{C}\leftarrow$ empty\;
\While{\emph{stack} not empty}{
    $u\leftarrow$ stack[end]; stack[end] = []\; 
    \If{\emph{visit}$[u]=0$}{
        comp $\leftarrow$ empty\;
        [visit, comp] $\leftarrow$ dfs($u$, $A^\top$, visit, comp)\;
        append sorted(comp) to $\mathcal{C}$\;
    }
}
\textbf{Procedure} dfs($u$, $A^\top$, visit, comp):\\ 
    visit$[u]\leftarrow1$; append $u$ to comp\;
    \For{$v\gets1$ \KwTo $n$}{
        \If{$A^{\top}_{uv}=1$ and \emph{visit}$[v]=0$}{
            [visit, comp] $\leftarrow$ dfs($v$, $A^\top$, visit, comp)\;
        }
    }

\textbf{Step 3: Identify Root SCCs}\;
$m \leftarrow$ length($\mathcal{C}$); root\_flags $\leftarrow [1,1,\dots,1]_{1\times m}$\;
\For{$i \gets 1$ \KwTo $m$}{
    \For{$j \gets 1$ \KwTo $m$}{
        \If{$i \neq j$}{
            \For{each $u \in \mathcal{C}_j$}{
                \For{each $v \in \mathcal{C}_i$}{
                    \If{$A_{uv}=1$}{
                        root\_flags[$i$] $\leftarrow 0$\;
                    }
                }
            }
        }
    }
}
$\mathcal{R} \leftarrow \{\,\mathcal{C}_i \;\text{whose}\; \text{root\_flags}[i] = 1\,\}$\;

\textbf{Output:} List of SCCs and list of root SCCs.
\end{algorithm}



Proposition 1 can be interpreted as the structural counterpart of the algebraic rank condition used in the $\mathcal{L}$ approach, where the Lie-derivative expansion of the outputs captures the same dependencies in an analytical form. Note that if the outputs $\bm{y}$ in \eqref{nonlinearsystem1new} depend on at least one state variable present in every root SCC of $\mathcal{D}$, the system is structurally observable. 



Let $\mathcal{X}_{\mathrm{r}}\subseteq\{x_{1},\dots,x_{n}\}$ denote the subset of state variables in root SCCs of $\mathcal{D}$, and $\mathcal{X}_{\mathrm{d}}=\{x_{1},\dots,x_{n}\}-\mathcal{X}_{\mathrm{r}}$ the set with possibly remaining state variables. Consider an output of the form $y=g(\mathcal{X}_{\mathrm{r}})$. A first-order Taylor series expansion of $g(\mathcal{X}_{\mathrm{r}})$ at $\bm{x}=\bm{x}_{0}$ is given by,
\vspace{-.2cm}
\begin{equation}
g(\mathcal{X}_{\mathrm{r}}) = g(\bm{x}_{0}) = \bm{c}^{\top}\bm{x}_{0} \label{eq.taylor-output}
\end{equation}

\vspace{-.2cm}
\noindent
where $\bm{c}$ denotes a vector of constants. Thus, from a structural standpoint, $y$ provides direct access to the state variables in $\mathcal{X}_{\mathrm{r}}$. Note that this is consistent with the $\mathcal{L}$ approach being defined locally, cf. Definition 3. Measuring any function of root–SCC variables injects information from all independent sources; this information is carried through the directed edges and reaches any remaining state variables in $\mathcal{X}_{\mathrm{d}}$. Algorithm \ref{algo:scc-mat} summarizes the process of identifying SCCs and root SCCs from an adjacency matrix built from the digraph. We are now able to present results on realistic models of power systems.
\vspace{-0.4cm}
\section{Numerical results}\label{sec:results}
We consider the Example 7.1 in \cite{Sauer1997}, a three-machine ($m\!=\!3$) nine-bus ($b\!=\!9$) system where each machine model,
\vspace{-.1cm}
\begin{align}
\dot{E}^{\prime}_{\text{q}i} &= c_{1i} \left(-E^{\prime}_{\text{q}i} - c_{2i}I_{\text{d}i} + E_{\text{fd}i} \right) \qquad i=1,\dots,m \label{eq:machine1}\\
\dot{E}^{\prime}_{\text{d}i} &= c_{3i} \left(-E^{\prime}_{\text{d}i} + c_{4i}I_{\text{q}i}\right) \nonumber\\
{\dot{\delta}_i} &= \omega_i - c_{5i} \nonumber\\
{\dot{\omega}_i} &= c_{6i} \left(T_{\text{M}i} - E^{\prime}_{\text{d}i} I_{\text{d}i} - E^{\prime}_{\text{q}i} I_{\text{q}i} - c_{7i}I_{\text{d}i} I_{\text{q}i} - c_{8i}(\omega_i - c_{5i}) \right) \nonumber\\
\dot{E}_{\text{fd}i} &= c_{9i} \left(-\left(c_{10i} + S_{\text{E}i}(E_{\text{fd}i})\right) E_{\text{fd}i} + V_{\text{R}i} \right) \nonumber\\
\dot{R}_{\text{f}i} &= c_{11i} \left(-R_{\text{f}i} + c_{12i}E_{\text{fd}i}\right) \nonumber\\
\dot{V}_{\text{R}i} &= c_{13i} \left(-V_{\text{R}i} + c_{14i}R_{\text{f}i} - c_{15i}E_{\text{fd}i} + c_{14i}(V_{\text{ref}i} - V_i) \right) \nonumber
\end{align}

\vspace{-.1cm}
\noindent
connects to the network via algebraic constraints, as follows. For each machine bus $i\!=\!1,\dots,m$, there are two stator voltage balance equations,
\vspace{-.1cm}
\begin{align}
E^{\prime}_{\text{d}i} - V_i \sin(\delta_i - \theta_i) - R_{\text{s}i} I_{\text{d}i} + X^{\prime}_{\text{q}i} I_{\text{q}i} &= 0 \label{eq:stator1} \\
E^{\prime}_{\text{q}i} - V_i \cos(\delta_i - \theta_i) - R_{\text{s}i} I_{\text{q}i} - X^{\prime}_{\text{d}i} I_{\text{d}i} &= 0 \label{eq:stator2}
\end{align}

\vspace{-.1cm}
\noindent
$V_ie^{j\theta_i} = V_i\cos\theta_i + jV_i\sin\theta_i = V_{\text{D}i} + jV_{\text{Q}i}$, and two network equations (presented below in polar form as a single equation)
\vspace{-.1cm}
\begin{align}
0 &= V_i e^{j\theta_i}(I_{\text{d}i} - j I_{\text{q}i}) e^{-j(\delta_i - \pi/2)} - \sum_{k=1}^n V_i V_k Y_{ik} e^{j(\theta_i - \theta_k - \alpha_{ik})} \nonumber \\
&\qquad + P_{\text{L}i} + jQ_{\text{L}i}. \label{eq:network}
\end{align}

\vspace{-.1cm}
For each load bus $i\!=\!m+1,\dots,b$, two network equations
\vspace{-.1cm}
\begin{align}
0= - \sum_{k=1}^n V_i V_k Y_{ik} e^{j(\theta_i - \theta_k - \alpha_{ik})} + P_{\text{L}i} + jQ_{\text{L}i}. \label{eq:network1}
\end{align}

\vspace{-.1cm}
\noindent
This model \eqref{eq:machine1}--\eqref{eq:network1} is well-known; see \cite{Sauer1997} for further details. Note that the graph approach is parameter-agnostic.

\begin{table}[!t]
\centering
\caption{Definition of constants in \eqref{eq:machine1}}
\vspace{-.3cm}
\begin{tabular}{lll}
\toprule
$c_{1i}=1/T^{\prime}_{\text{d0}i}$ & $c_{6i}=\omega_s/(2H_i)$ & $c_{11i}=1/T_{\text{F}i}$ \\
$c_{2i}=X_{\text{d}i}-X^{\prime}_{\text{d}i}$ & $c_{7i}=X^{\prime}_{\text{q}i}-X^{\prime}_{\text{d}i}$ & $c_{12i}=K_{\text{F}i}/T_{\text{F}i}$ \\
$c_{3i}=1/T^{\prime}_{\text{q0}i}$ & $c_{8i}=D_{i}$ & $c_{13i}=1/T_{\text{A}i}$ \\
$c_{4i}=X_{\text{q}i} - X^{\prime}_{\text{q}i}$ & $c_{9i}=1/T_{\text{E}i}$ & $c_{14i}=K_{\text{A}i}$ \\
$c_{5i}=\omega_s$ & $c_{10i}=K_{\text{E}i}$ & $c_{15i}=K_{\text{A}i}K_{\text{F}i}/T_{\text{F}i}$ \\
\bottomrule
\end{tabular}
\label{tab:my_label}
\end{table}



\vspace{-.3cm}
\subsection*{Case 1: Decentralized dynamic state estimation}
Following \cite{Singh2014}, we treat voltages as inputs and currents as outputs to separate the dynamics of each generator from the rest of the network without ignoring their interaction. Thus,
\vspace{-.1cm}
\begin{equation}\label{vectorscentralized}
\begin{aligned}
\bm{x} &= [\, E^{\prime}_{\text{q}i}\;\; E^{\prime}_{\text{d}i}\;\; \delta_i\;\; \omega_i\;\; E_{\text{fd}i}\;\; R_{\text{f}i}\;\; V_{\text{R}i} \,]^\top, \\
\bm{u} &= [\, T_{\text{M}i}\;\; V_{\textnormal{ref}i}\;\; V_{\text{D}i}\;\; V_{\text{Q}i} \,]^\top, 
\;\text{and}\;\;
\bm{y} = [\, I_{\text{D}i}\;\; I_{\text{Q}i} \,]^\top,
\end{aligned}
\end{equation}

\vspace{-.1cm}
\noindent
where
\vspace{-.2cm}
\begin{equation}
I_{\text{D}i}+jI_{\text{Q}i} = \left( I_{\text{d}i}+jI_{\text{q}i} \right)e^{j\left(\delta_{i}-\frac{\pi}{2}\right)} \label{eq.IDIQ}    
\end{equation}

\vspace{-.1cm}
\noindent
This particular choice of $\{V_{\text{D}i}, V_{\text{Q}i}, I_{\text{D}i}, I_{\text{Q}i}\}$ is consistent with the availability of synchrophasor measurements at the terminal of each machine. Now, following \cite{Sauer1997}, we define
\vspace{-.1cm}
\begin{equation}
\bm{Z}_{\text{dq}i} \triangleq
\begin{bmatrix}
R_{\text{s}i} & -X^{\prime}_{\text{q}i} \\
X^{\prime}_{\text{d}i} & R_{\text{s}i}
\end{bmatrix}. \label{eq.Z}
\end{equation}

\vspace{-.1cm}
Using \eqref{eq.Z} to algebraically manipulate \eqref{eq:stator1}--\eqref{eq:stator2} yields
\vspace{-.1cm}
\begin{equation}
\begin{bmatrix}
I_{\text{d}i} \\[0pt]
I_{\text{q}i}
\end{bmatrix}
=
[\bm{Z}_{\text{dq}i}]^{-1}
\begin{bmatrix}
E^{\prime}_{\text{d}i} - V_i \sin(\delta_i - \theta_i) \\[0pt]
E^{\prime}_{\text{q}i} - V_i \cos(\delta_i - \theta_i)
\end{bmatrix},\; i\!=\!1,\dots,m. \label{eq.IdIq}
\end{equation}

\vspace{-.1cm}
\noindent
From \eqref{eq.IDIQ} and \eqref{eq.IdIq}, it is evident that $I_{\text{D}i}$ and $I_{\text{Q}i}$ are functions of the state variables $\{E^{\prime}_{\text{q}i},E^{\prime}_{\text{d}i},\delta\}$. Next, because the graph approach requires only knowledge of the dependencies in a state-space model, we will use the simplified notation
\vspace{-.15cm}
\begin{align}\label{eq.algebraic}
I_{\text{D}i} &= h_{1} \left(\{\mathcal{X}\}, \{\mathcal{U}\} \right) = h_{1} \left( \{E_{\text{q}i}^{\prime},\,E_{\text{d}i}^{\prime},\,\delta_i\}, \{V_{\text{D}i},\,V_{\text{Q}i}\} \right) \nonumber\\
I_{\text{Q}i} &= h_{2} \left( \{E_{\text{q}i}^{\prime},\,E_{\text{d}i}^{\prime},\,\delta_i\},\{V_{\text{D}i},\,V_{\text{Q}i}\} \right) 
\end{align}

\vspace{-.1cm}
\noindent
where $\mathcal{X}$ (respectively, $\mathcal{U}$) denotes a subset of the variables in $\bm{x}$ (respectively, $\bm{u}$), cf. \eqref{nonlinearsystem1new}. Substituting $I_{\text{D}i}$ and $I_{\text{Q}i}$ in \eqref{eq:machine1},
\vspace{-.15cm}
\begin{align}\label{eq:decentralized_machine}
\dot{E}_{\text{q}i}^{\prime} &= f_{1} \left( \{E_{\text{q}i}^{\prime},\,E_{\text{d}i}^{\prime},\,\delta_i,\,E_{\text{fd}i}\},\, \{V_{\text{D}i},\,V_{\text{Q}i}\} \right) \nonumber\\
\dot{E}_{\text{d}i}^{\prime} &= f_{2} \left( \{E_{\text{q}i}^{\prime},\,E_{\text{d}i}^{\prime},\,\delta_i\},\, \{V_{\text{D}i},\,V_{\text{Q}i}\}\right) \nonumber\\
\dot{\delta_i} &= f_{3} \left( \{\omega_i\},\, \{\} \right) {} \nonumber\\
\dot{\omega_i} &= f_{4} \left( \{E_{\text{q}i}^{\prime},\,E_{\text{d}i}^{\prime},\,\delta_i,\,\omega_i\},\, \{T_{\text{M}i}\} \right) \nonumber\\
\dot{E}_{\text{fd}i} &= f_{5} \left( \{E_{\text{fd}i},\,V_{\text{R}i}\},\, \{\} \right) \nonumber\\
\dot{R}_{\text{f}i} &= f_{6} \left( \{E_{\text{fd}i},\,R_{\text{f}i}\},\, \{\} \right) \nonumber\\
\dot{V}_{\text{R}i} &= f_{7} \left( \{E_{\text{fd}i},\,R_{\text{f}i},\,V_{\text{R}i}\},\, \{V_{\textnormal{ref}i}\} \right).
\end{align}

\vspace{-.1cm}
\noindent
From \eqref{eq:decentralized_machine}, the adjacency matrix
\vspace{-.1cm}
\begin{equation}
\setlength{\arraycolsep}{7pt}
\renewcommand{\arraystretch}{.7}
\bm{A} = \left[\begin{array}{ccccccc}
1& 1& 1& & 1& & \\
1& 1& 1& & & & \\
 & & & 1& & & \\
1& 1& 1& 1& & & \\
 & & & & 1& & 1\\
 & & & & 1& 1& \\
 & & & & 1& 1& 1\\
\end{array}\right], \label{eq.adj}
\end{equation}

\begin{figure}[!t]\centering
\begin{tikzpicture}[node distance={15mm}, thick, main/.style = {draw, circle}] 
\node[main] (90) [              color=white] {}; 
\node[main] (91) [below of= 90, color=white] {}; 
\filldraw[densely dotted, color=red!60,  fill=red!15,  very thick, opacity=0.4] (-3.85,-2.35) rectangle (.7,.85);
\filldraw[densely dotted, color=blue!60, fill=blue!15, very thick, opacity=0.4] (-8.0,-2.35) rectangle (-5.3,.85);
\draw (-0.9, 0.8) node[anchor=north west] {\small root SCC};
\draw (-8.0,-1.8) node[anchor=north west] {\small SCC};
\node[main] (1)  [left of= 90, fill=white] {$E^{\prime}_{\text{q}}$};
\node[main] (2)  [left of= 91, fill=white] {$E^{\prime}_{\text{d}}$};
\node[main] (3)  [right of=90, fill=white] {$V_{\text{D}}$};
\node[main] (4)  [right of=91, fill=white] {$V_{\text{Q}}$};
\node[main] (5)  [ left of= 2, fill=white] {$\delta$};
\node[main] (6)  [ left of= 1, fill=white] {$\omega$};
\node[main] (7)  [ left of= 6, fill=white] {$T_{\text{M}}$};
\node[main] (8)  [ left of= 7, fill=white] {$E_{\text{fd}}$};
\node[main] (9)  [below of= 8, fill=white] {$V_{\text{R}}$};
\node[main] (10) [right of= 9, fill=white] {$V_{\text{ref}}$};
\node[main] (11) [ left of= 8, fill=white] {$R_{\text{f}}$};
\draw[->] (1)  to [out= 115,in= 65,looseness=3] (1);
\draw[->] (2)  to [out=-115,in=-65,looseness=3] (2);
\draw[->] (6)  to [out= 115,in= 65,looseness=3.5] (6);
\draw[->] (8)  to [out= 135,in= 95,looseness=3] (8);
\draw[->] (9)  to [out=-115,in=-65,looseness=3.5] (9);
\draw[->] (11)  to [out= 115,in= 65,looseness=3.5] (11);
\draw[->] (1) to [out= -75,in=  75,looseness=0.9] (2);
\draw[->] (1) to [out= 140,in=  40,looseness=0.6] (8);
\draw[->] (2) to [out= 105,in=-105,looseness=0.9] (1);
\draw[->] (2) to [out= -30,in= -45,looseness=1.8] (3);
\draw[<-] (5) to [out= 105,in=-105,looseness=0.9] (6);
\draw[<-] (6) to [out= -75,in=  75,looseness=0.9] (5);
\draw[->] (6) to [out=-130,in=-145,looseness=2.3] (2);
\draw[->] (8) to [out= -75,in=  75,looseness=0.9] (9);
\draw[->] (9) to [out= 105,in=-105,looseness=0.9] (8);
\draw[->] [color=black]  (1) --  (3);
\draw[->] [color=black]  (1) --  (4);
\draw[->] [color=black]  (1) --  (5);
\draw[->] [color=black]  (2) --  (4);
\draw[->] [color=black]  (2) --  (5);
\draw[->] [color=black]  (6) --  (1);
\draw[->] [color=black]  (6) --  (7);
\draw[->] [color=black]  (9) -- (10);
\draw[->] [color=black]  (9) -- (11);
\draw[->] [color=black] (11) --  (8);
\end{tikzpicture}
\vspace{-.6cm}
\caption{Digraph for Case 1: Decentralized dynamic state estimation.}
\label{fig:inference_diagram}
\end{figure}
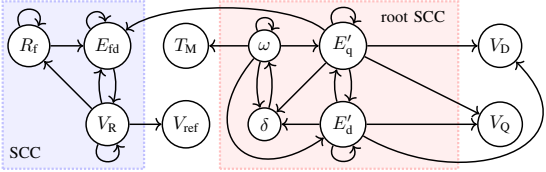

\vspace{-.1cm}
\noindent
where each row is associated with a differential equation $\dot{x}_i$, in the order they appear in \eqref{eq:decentralized_machine}, and each column with a state variable $x_i$. An entry $a_{ij}$ in \eqref{eq.adj} is assigned the value `1' if the differential equation includes $x_j$ (i.e., $\frac{\partial \dot{x}_i}{\partial x_j} \neq 0$), and zero otherwise; zeros are omitted in \eqref{eq.adj} for clarity. We apply a modified Kosaraju-Sharir's Algorithm \ref{algo:scc-mat} \cite{SHARIR198167} to the adjacency matrix to identify strongly connected components (SCCs) and systematically determine the root SCCs. Fig. \ref{fig:inference_diagram} depicts a digraph constructed from \eqref{eq.adj}. The exciter variables $R_{\text{f}i}$, $V_{\text{R}i}$, and $E_{\text{fd}i}$ form an SCC, but not a root SCC, since there is an incoming edge. Conversely, $E_{\text{q}i}^\prime$, $E_{\text{d}i}^\prime$, $\omega_i$ and $\delta_i$ form a root SCC. A measurement depending on \textit{at least} one variable in the root SCC makes the system structurally observable, cf. \eqref{eq.taylor-output}. Clearly, $I_{\text{D}i}$ and $I_{\text{Q}i}$ depend on three variables within the root SCC, satisfying the condition for structural observability.

We verify this result using the $\mathcal{L}$ approach. We construct an observability matrix by computing successive Lie derivatives of the output $\bm{y}=[I_{\text{D}i}\; I_{\text{Q}i}]^\top$. The $\mathcal{L}$ approach yields a full-rank observability matrix of dimension~7, confirming that the system is observable in the sense of Definition 3. Table \ref{tab:dse-time} compares the average computation time required to check observability using the $\mathcal{L}$ and graph approaches.

\vspace{-.3cm}
\subsection*{Case 2: Centralized DSE}
We perform a Kron reduction \cite{Dorfler2013} of \eqref{eq:machine1}--\eqref{eq:network} to eliminate its algebraic constraints. This yields $\bm{x}=\left[\bm{x}_{1}^{\top}\,\, \bm{x}_{2}^{\top}\,\, \bm{x}_{3}^{\top}\right]^{\top}$, $\bm{u}=\left[\bm{u}_{1}^{\top}\,\, \bm{u}_{2}^{\top}\,\, \bm{u}_{3}^{\top}\right]^{\top}$, and $\bm{y}=\left[\bm{y}_{1}^{\top}\,\, \bm{y}_{2}^{\top}\,\, \bm{y}_{3}^{\top}\right]^{\top}$, where 
\vspace{-.1cm}
\begin{equation*}
\begin{aligned}
\bm{x}_{i} &= [\, E^{\prime}_{\text{q}i}\;\; E^{\prime}_{\text{d}i}\;\; \delta_i\;\; \omega_i\;\; E_{\text{fd}i}\;\; R_{\text{f}i}\;\; V_{\text{R}i} \,]^\top, \\
\bm{u}_{i} &= [\, T_{\text{M}i}\;\; V_{\textnormal{ref}i} \,]^\top, 
\;\text{and}\;\;
\bm{y}_{i} = [\, V_{\text{D}i}\;\; V_{\text{Q}i} \,]^\top.
\end{aligned}
\end{equation*}

\vspace{-.1cm}
The updated model is written in the simplified form:
\vspace{-.1cm}
\begin{align}\label{eq:centralized_3machine}
\dot{E}_{\text{q}i}^{\prime} &= f_{1i}^{\prime} \left( \{E_{\text{q}j}^{\prime},\,E_{\text{d}j}^{\prime},\,\delta_j,\,E_{\text{fd}i} \}_{j=1}^{3},\, \{\}\right) \\
\dot{E}_{\text{d}i}^{\prime} &= f_{2i}^{\prime} \left( \{E_{\text{q}j}^{\prime},\,E_{\text{d}j}^{\prime},\,\delta_j\}_{j=1}^{3},\, \{\} \right) \nonumber\\
\dot{\delta}_{i} &= f_{3i}^{\prime} \left( \{\omega_i\},\, \{\} \right) \nonumber\\
\dot{\omega}_{i} &= f_{4i}^{\prime} \left( \{E_{\text{q}j}^{\prime},\,E_{\text{d}j}^{\prime},\,\delta_j,\,{\omega_i}\}_{j=1}^{3},\, \{T_{\text{M}i}\} \right) \nonumber\\
\dot{E}_{\text{fd}i} &= f_{5i}^{\prime} \left( \{E_{\text{fd}i},\,V_{\text{R}i}\},\, \{\} \right) \nonumber\\
\dot{R}_{\text{f}i} &= f_{6i}^{\prime} \left( \{E_{\text{fd}i},\,R_{\text{f}i}\},\, \{\} \right) \nonumber\\
\dot{V}_{\text{R}i} &= f_{7i}^{\prime} \left( \{E_{\text{fd}i},\,R_{\text{f}i},\, V_{\text{R}i}\}, \{V_{\text{ref}i}\} \right), \qquad i=\{1,2,3\}. \nonumber
\end{align}

\vspace{-.1cm}
A digraph is less amenable to visualization in this case due to the strong coupling among state variables. Therefore, we only show the adjacency matrix, see Table \ref{tab:incidence_matrix}. Algorithm \ref{algo:scc-mat} identifies three SCCs, $\{V_{\text{R}1}, R_{\text{f}1}, E_{\text{fd}1}\}$, $\{V_{\text{R}2}, R_{\text{f}2}, E_{\text{fd}2}\}$, and $\{V_{\text{R}3}, R_{\text{f}3}, E_{\text{fd}3}\}$;  and one root SCC, $\{\dot{E}_{\text{q}1}^{\prime}, \dot{E}_{\text{d}1}^{\prime}, \delta_1, \omega_1, \dot{E}_{\text{q}2}^{\prime},$ $\dot{E}_{\text{d}2}^{\prime},\delta_2, \omega_2, \dot{E}_{\text{q}3}^{\prime}, \dot{E}_{\text{d}3}^{\prime}, \delta_3, \omega_3\}$. Measuring at least one variable in each root SCC yields a structurally observable system. For instance, measuring
\vspace{-.1cm}
\begin{equation*}
V_{\text{D}i}= g_3\left(E_{\text{q}i}^{\prime},\,E_{\text{d}i}^{\prime},\,\delta_i\right)  \;\;\text{and}\;\;
V_{\text{Q}i}= g_{4} \left(E_{\text{q}i}^{\prime},\,E_{\text{d}i}^{\prime},\,\delta_i \right),
\end{equation*}

\vspace{-.1cm}
\noindent
$i=\{1,2,3\}$, yields structural observability. The $\mathcal{L}$ approach returns a full $\text{rank}(\bm{O})=21$ (3 generators, 7 state variables per generator) for the same measurement setup and, on average, requires 2 hours to compute. The average computation time of the graph approach is less than 5 seconds, an average $\texttt{1440}\times$ speedup even in such a small test system. See Table \ref{tab:dse-time}.

\section{Conclusions and future work} \label{sec:conclusion}
Current methods for observability analysis do not scale, especially in \emph{centralized} dynamic state estimation. Even in the small test system we examined, with only three generators and twenty-one state variables, our graph-based approach yields a substantial reduction in computation time. The proposed method will facilitate new advances in power system dynamic state estimation, including new optimization formulations for synchrophasor measurement placement. Future work will focus on structure-preserving models of power system networks and incorporate models of inverter-based resources.

\begin{table}[!t]
\centering
\caption{Average execution times for Cases 1 and 2.}
\vspace{-.3cm}
\label{tab:dse-time}
\footnotesize
\def\arraystretch{.1} 
\begin{tabular}{l c c}
\toprule
\textbf{Approach} & \textbf{Case 1: Decentralized DSE} & \textbf{Case 2: Centralized DSE} \\
\midrule
$\mathcal{L}$ & $<$ 5 minutes & 2 hours \\
\midrule
Graph & $\sim$ 1 second & $<$ 5 seconds \\
\bottomrule
\end{tabular}
\end{table}

\begin{table}[!t]
\centering
\caption{Adjacency matrix for the centralized DSE case}
\label{tab:incidence_matrix}
\tiny
\setlength{\arraycolsep}{2pt}
\renewcommand{\arraystretch}{1.2}
\vspace{-.6cm}
\[
\begin{array}{c|*{21}{c}}
 & \rotatebox{90}{$E^{\prime}_{\text{d1}}$} 
 & \rotatebox{90}{$E^{\prime}_{\text{d2}}$}
 & \rotatebox{90}{$E^{\prime}_{\text{d3}}$} 
 & \rotatebox{90}{$E^{\prime}_{\text{q1}}$}
 & \rotatebox{90}{$E^{\prime}_{\text{q2}}$}
 & \rotatebox{90}{$E^{\prime}_{\text{q3}}$}
 & \rotatebox{90}{$\delta_{1}$} 
 & \rotatebox{90}{$\delta_{2}$}
 & \rotatebox{90}{$\delta_{3}$}
 & \rotatebox{90}{$\omega_{1}$}
 & \rotatebox{90}{$\omega_{2}$}
 & \rotatebox{90}{$\omega_{3}$}
 & \rotatebox{90}{$E_{\text{fd1}}$}
 & \rotatebox{90}{$E_{\text{fd2}}$}
 & \rotatebox{90}{$E_{\text{fd3}}$}
 & \rotatebox{90}{$R_{\text{f1}}$}
 & \rotatebox{90}{$R_{\text{f2}}$}
 & \rotatebox{90}{$R_{\text{f3}}$}
 & \rotatebox{90}{$V_{\text{R1}}$}
 & \rotatebox{90}{$V_{\text{R2}}$}
 & \rotatebox{90}{$V_{\text{R3}}$} \\\hline
\dot{E}^{\prime}_{\text{d1}} & 1 & 1 & 1 & 1 & 1 & 1 & 1 & 1 & 1 & & & & & & & & & & & & \\
\dot{E}^{\prime}_{\text{d2}} & 1 & 1 & 1 & 1 & 1 & 1 & 1 & 1 & 1 & & & & & & & & & & & & \\
\dot{E}^{\prime}_{\text{d3}} & 1 & 1 & 1 & 1 & 1 & 1 & 1 & 1 & 1 & & & & & & & & & & & & \\
\dot{E}^{\prime}_{\text{q1}} & 1 & 1 & 1 & 1 & 1 & 1 & 1 & 1 & 1 & & & & 1 & & & & & & & & \\
\dot{E}^{\prime}_{\text{q2}} & 1 & 1 & 1 & 1 & 1 & 1 & 1 & 1 & 1 & & & & & 1 & & & & & & & \\
\dot{E}^{\prime}_{\text{q3}} & 1 & 1 & 1 & 1 & 1 & 1 & 1 & 1 & 1 & & & & & & 1 & & & & & & \\
\dot{\delta}_{1} & & & & & & & & & & 1 & & & & & & & & & & & \\
\dot{\delta}_{2} & & & & & & & & & & & 1 & & & & & & & & & & \\
\dot{\delta}_{3} & & & & & & & & & & & & 1 & & & & & & & & & \\
\dot{\omega}_{1} & 1 & 1 & 1 & 1 & 1 & 1 & 1 & 1 & 1 & 1 & & & & & & & & & & & \\
\dot{\omega}_{2} & 1 & 1 & 1 & 1 & 1 & 1 & 1 & 1 & 1 & & 1 & & & & & & & & & & \\
\dot{\omega}_{3} & 1 & 1 & 1 & 1 & 1 & 1 & 1 & 1 & 1 & & & 1 & & & & & & & & & \\
\dot{E}_{\text{fd1}} & & & & & & & & & & & & & 1 & & & 1 & & & & & \\
\dot{E}_{\text{fd2}} & & & & & & & & & & & & & & 1 & & & 1 & & & & \\
\dot{E}_{\text{fd3}} & & & & & & & & & & & & & & & 1 & & & 1 & & & \\
\dot{R}_{\text{f1}} & & & & & & & & & & & & & 1 & & & 1 & & & & & \\
\dot{R}_{\text{f2}} & & & & & & & & & & & & & & 1 & & & 1 & & & & \\
\dot{R}_{\text{f3}} & & & & & & & & & & & & & & & 1 & & & 1 & & & \\
\dot{V}_{\text{R1}} & & & & & & & & & & & & & 1 & & & 1 & & & 1 & & \\
\dot{V}_{\text{R2}} & & & & & & & & & & & & & & 1 & & & 1 & & & 1 & \\
\dot{V}_{\text{R3}} & & & & & & & & & & & & & & & 1 & & & 1 & & & 1 \\
\end{array}
\]
\end{table}

\bibliographystyle{IEEEtran}
\bibliography{lib}

\end{document}